\documentclass[12pt,a4paper]{article}
\usepackage{graphics}
\usepackage{cite,a4wide}
\newcommand{\beq}{\begin{equation}}
\newcommand{\eeq}{\end{equation}}
\newcommand{\bea}{\begin{eqnarray}}
\newcommand{\eea}{\end{eqnarray}}
\renewcommand{\d}{\delta}

\renewcommand{\b}{\beta}
\renewcommand{\a}{\alpha}
\renewcommand{\ni}{\noindent}

\newcommand{\m}{\mu}

\newcommand{\s}{\sigma}

\newcommand{\vn}{\vec{n}}
\newcommand{\vL}{\vec{L}}
\newcommand{\va}{\vec{\alpha}}
\newcommand{\vH}{\vec{H}}
\newcommand{\W}{{\cal G}}

\newcommand{\oh}{\frac{1}{2}}

\newcommand{\dg}{\dagger}
\newcommand{\non}{\nonumber}

\newcommand{\rf}[1]{(\ref{#1})}
\newcommand{\ra}{\rightarrow}

\begin{document}

\hfill October 1997

\begin{center}

\vspace{32pt}

  {\Large \bf Casimir Scaling from Center Vortices:} 

\bigskip

  { \bf Towards an Understanding of the Adjoint String Tension}

\end{center}

\vspace{18pt}

\begin{center}
{\sl M. Faber${}^a$, J. Greensite${}^b$,
and {\v S}. Olejn\'{\i}k${}^c$}

\end{center}

\vspace{18pt}

\begin{tabbing}

{}~~~~~~~~~~~~~~~~~~~~\= blah  \kill
\> ${}^a$ Inst. f\"ur Kernphysik, Technische Universit\"at Wien, \\
\> ~~A-1040 Vienna, Austria.  E-mail: {\tt faber@kph.tuwien.ac.at} \\
\\
\> ${}^b$ The Niels Bohr Institute, DK-2100 Copenhagen \O, \\
\> ~~Denmark.  E-mail: {\tt greensite@nbivms.nbi.dk} \\
\\
\> ${}^c$ Institute of Physics, Slovak Academy of Sciences, \\
\> ~~SK-842 28 Bratislava, Slovakia.  E-mail: {\tt fyziolej@savba.sk}

\end{tabbing}

\vspace{18pt}

\begin{center}

{\bf Abstract}

\end{center}

\bigskip

   We argue that the approximate ``Casimir scaling'' of the string
tensions of higher-representation Wilson loops is an effect due to 
the finite thickness of center vortex configurations. It is shown, in 
the context of a simple model of the $Z_2$ vortex core, how vortex 
condensation in Yang-Mills theory can account for both Casimir scaling 
in intermediate size loops, and color-screening in larger loops. An 
implication of our model is that the deviations from exact Casimir scaling, 
which tend to grow with loop size, become much more pronounced as the 
dimensionality of the group representation increases.

\vfill

\newpage

\section{Introduction}

   There is increasing numerical evidence \cite{PRD97,Zako,TK,LR}
supporting the Center Vortex Theory of quark confinement 
\cite{tHooft,Mack,Cop,Corn1,Feyn}, which was put forward in the late 1970's.
Briefly, a center vortex is a topological field configuration
which is line-like (in D=3 dimensions) or surface-like (in D=4 dimensions) 
having some finite thickness.  Creation of a center vortex can be
regarded, outside the line-like or surface-like ``core,'' as a discontinuous
gauge transformation of the background, with a discontinuity associated with 
the gauge group center.  Creation of a center vortex linked to
a Wilson loop, in the fundamental representation of $SU(N)$, has the 
effect of multiplying the Wilson loop by an element of the gauge group
center, i.e.
\beq
       W(C) \ra e^{i2\pi n /N} W(C) ~~~~~~~~~~~~~ n=1,2,...,N-1
\label{vortex}
\eeq
The vortex theory, in essence, states that the area law for Wilson
loops is due to quantum fluctuations in the number of center vortices
linking the loop.

   Paradoxically, this emphasis on the center of the gauge group can
be viewed both as a vital strength of the theory, and also as a fatal 
weakness. Both aspects are apparent when we consider the force between static
quarks in an $SU(N)$ gauge theory, whose color charge lies in the adjoint 
representation.  The QCD vacuum will not tolerate a linear potential
between adjoint quarks over an infinite range; this is simply because
adjoint color charges can be screened by gluons.  Asymptotically, the force
between adjoint quarks must drop to zero, and this is exactly what happens
in the center vortex theory.  The adjoint representation
transforms trivially under the group center; adjoint
Wilson loops are unaffected by center vortices, unless the core of
the vortex happens to overlap the perimeter of the loop.  As a result,
large loops have only perimeter falloff, and the force between 
adjoint quarks vanishes asymptotically.  The argument extends to any color
representation which transforms trivially under the $Z_N$ center of the
gauge group.  The fact that center vortices make such a clear distinction 
between those color charges which \emph{should} be confined, and those charges
which should not, is one of the most attractive features of the theory.

   The fatal weakness aspect was first pointed out in ref. \cite{JGH}.  
Consider the large-$N$ limit, which has the factorization property 
$<AB>=<A><B>$, where $A$ and $B$ are any two gauge-invariant operators.  Then 
\beq
       <W_A(C)> = <W_F(C) W_F^{\dg}(C)> = <W_F(C)>^2
\eeq
where $W_{A,F}(C)$ denotes, respectively, Wilson loops in the adjoint 
and fundamental representations.  An immediate consequence is that
confinement of fundamental representation quarks implies confinement
in the adjoint representation, with string tension $\s_A = 2\s_F$.
This is possible because color screening by gluons is a $1/N^2$ suppressed
process, so at large $N$ the vacuum \emph{can} support an adjoint
string.  But adjoint loops are insensitive to center vortices, as noted
above.  The apparent conclusion is that center vortices cannot be the 
confinement mechanism at large $N$.
    
   Even more troubling is the fact that the existence of an adjoint string 
tension is not peculiar to large-$N$.
Many numerical experiments in $SU(2)$ and $SU(3)$ lattice gauge theory
have shown that flux-tubes form, and a linear potential is established,
between quarks in the adjoint (and higher) representations \cite{Cas}.  
The string tension is representation-dependent, and appears to be
roughly proportional to the eigenvalue of the quadratic Casimir operator
of the representation.  Thus, for an $SU(2)$ gauge theory
\beq
       \s_j \approx {4\over 3} j(j+1) \s_{1/2}
\eeq
where $\s_j$ is the string tension in representation $j$.
The region where this relation is valid, from the onset of confinement
to the onset of color-screening, we call the ``Casimir-scaling regime''
\cite{lat96,Cas1}.  Of course, the color charge of higher-representation 
quarks is eventually
screened by gluons, and the force between quarks then depends
only on the transformation properties of the representation
with respect to the gauge group 
center; i.e. on the ``n-ality'' of the representation. 
Asymptotically, for an $SU(2)$ gauge group,
\beq
        \s_j = \left\{ \begin{array}{cl}
                          \s_{1/2} & j = \mbox{half-integer} \\
                              0    & j = \mbox{integer} \end{array} \right.
\eeq
Color screening, although it must occur for adjoint quarks at sufficiently
large separation, is very difficult to observe in numerical 
simulations.\footnote{At least, this is difficult at zero temperature.  
Color screening \emph{has} been observed in certain finite-temperature 
studies, cf. M\"{u}ller et al. in ref. \cite{Cas}.}
Existing Monte Carlo studies of the QCD string have mainly probed the 
Casimir scaling regime.  

   In short, there is a linear potential between adjoint quarks in some 
finite range of distances, with approximate Casimir scaling of the string 
tensions, even at $N=2$.  Casimir scaling should become exact, in a region 
extending from the confinement scale to infinity, in the 
$N \ra \infty$ limit.  Yet according to the center vortex theory, it would
appear that string formation between adjoint quarks is impossible, at any $N$,
at any distance scale.  This has always seemed to us a good reason for 
discarding the vortex theory.
   
   But suppose $-$ and in our view the numerical evidence 
\cite{PRD97,Zako,TK,LR} is
becoming persuasive $-$ that center vortices really \emph{are} the
confining configurations, at least  for quarks in the fundamental 
representation.  Then either there is some other mechanism for
inducing a linear potential between adjoint quarks, or else there must be a
loophole in the ``fatal weakness'' arguments.  The first alternative
does not seem very economic, and in any case we have no insight, at present,
in that direction.  We will concentrate instead on the second
possibility because there is, in fact, one possible loophole.
 
\section{Thick Vortices and the Loop Perimeter}

   The statement that adjoint loops are unaffected by center vortices
contains one slight caveat: They are unaffected \emph{unless} the
vortex core somewhere overlaps the perimeter of the loop.  At first
sight this caveat seems irrelevant; for large loops the effect can
only contribute to the perimeter falloff. But suppose that the vortex
thickness is actually quite large: on the order of, and perhaps
exceeding, the typical diameters of low-lying hadrons.  What would be
the effect of vortices on Wilson loops whose area is smaller than, or
comparable to, the vortex cross section? We will study this question
in the context of a simple model of vortex/perimeter overlaps, mainly
in $SU(2)$ lattice gauge theory.

   In D=3 dimensions, or on a constant-time hypersurface in D=4 dimensions, 
a vortex is a closed tube of magnetic flux.  For simplicity we consider 
planar and, in D=4 dimensions, spacelike Wilson loops.
If the vortex is linked to a Wilson loop, with winding number $=1$,
then the vortex pierces the minimal area of the loop an odd number 
of times.  If the vortex is not linked to the loop, it
either does not pierce the minimal area at all, or pierces
it an even number of times.  If, for the moment, we ignore
the finite radii of the vortex tubes, then the effect of vortices
on a Wilson loop
is simple: For every instance where the minimal surface is pierced by a 
center vortex, insert a center element $-I$ somewhere along the loop, 
i.e.
\beq
     W(C) = \mbox{Tr}[UU....U] \ra \mbox{Tr}[UU..(-I)..U]
\label{minus}
\eeq
In principle we should place the $-I$ at the point of discontinuity
of the gauge transformation which creates the vortex.  However, since
$-I$ commutes with everything, the placement is arbitrary; and this is
related to the fact that the Dirac sheet of a center vortex can
be moved about by gauge transformations.

   Denote by $f$ the probability that any given plaquette on the
lattice is ``pierced'' by a vortex; i.e. a line running through the center
of the vortex tube intersects the plaquette.  The area law for Wilson
loops is then trivially derived from the assumption that these probabilities,
for plaquettes in a plane, are uncorrelated.\footnote{Some related ideas are
found in ref. \cite{Poli}.}  In that case, one has
\bea
       <W(C)> &=& \prod_{x\in A} \{ (1-f) + f(-1) \} <W_0(C)>
\non \\
        &=& \exp[-\s(C) A] <W_0(C)>
\eea
where the string tension is
\beq
       \s = - \ln(1-2f)
\eeq
and where $<W_0(C)>$ is the expectation value of the loop with the constraint
that no vortices pierce the minimal area.  The quantity $<W_0(C)>$ can be 
(and has been) computed from lattice Monte Carlo, cf. ref. 
\cite{PRD97,Zako}.  In those computations, it is found that 
$W_0(C)$ does not have an area law falloff.

   By the same argument, the string tension for loops in any  $j=$ 
half-integer representation is the same as for $j=1/2$, while the
string tension for $j=$ integer vanishes.
Of course the argument is too simple in a number of respects,
e.g. there is likely to be some short-range correlation between the
$f$ probabilities of nearby plaquettes.  This point, however, is not crucial 
to the discussion.  What is more important is that we have ignored the
finite radii of the vortices.  Equation \rf{minus} is only true if
the core of the vortex, where it crosses the plane of the loop, is
entirely contained in the minimal area of the loop.  If the core overlaps
the perimeter of the loop, then eq. \rf{minus} cannot be quite right.

   What is needed is a full-fledged theory of center vortices, perhaps
something along the lines of the old Copenhagen vacuum \cite{Cop}, which 
would explain how eq. \rf{minus} should be modified when the vortex core is
not entirely enclosed within the loop.  In lieu of that,
we will just consider a simplified picture in which the center element $-I$
in \rf{minus} is replaced by a group element $G$, which interpolates
smoothly from $-I$, if the core is contained entirely with the loop, 
to $+I$, if the core is entirely exterior.  Our assumption, for Wilson
loops in any group representation $j$, is the following:  

\begin{description}
\item
\ni $\underline{\mbox{\bf Assumption 1}}$ 


  The effect of creating a center vortex piercing the minimal area of a 
Wilson loop may be represented by the insertion of 
a unitary matrix $G$ at some point along the loop
\beq
     W(C) = \mbox{Tr}[UU....U] \ra \mbox{Tr}[UU..G..U]
\eeq
where
\bea
        G(x,S) &=& \exp[i\a_C(x)\vn \cdot \vL]
\non \\
                 &=& S \exp[i\a_C(x) L_3] S^\dg
\eea
The $L_i$ are group generators in representation $j$, 
$\vn$ is a unit 3-vector, and $S$ is an $SU(2)$ group element in
the $j$-representation.
\end{description}

   The parameter $\a_C(x) \in [0,2\pi]$ depends on what fraction of the
vortex core is enclosed by the loop; thus it depends on both the shape 
of the loop $C$, and the position $\vec{x}$ of the center of the vortex core,
relative to the perimeter, in the plane of the loop. 
If the core is entirely enclosed by
the loop, then $\a_C(x)=2\pi$, conversely, if the core is entirely outside
the minimal area of the loop, then $\a_C(x)=0$.   In an abelian theory,
this first assumption would be completely correct, where Tr[$G$] would be the
value of the loop for a vortex created on a classical vacuum background.
In a non-abelian theory the assumption might be quantitatively correct for 
expectation values (i.e. averaging over group orientations $S$ and over 
small quantum fluctuations $U_\m$ around the vortex background); this would 
be quite sufficient for our purposes.

   Generalizing a little further,
if we create some number $m$ of vortices in the loop, centered at
positions $x_1,x_2,...,x_m$, then
\bea
      W(C) &\ra&  W[C;\{x_i,S_i\}] 
\non \\
 &=& \mbox{Tr}[U..UG(x_a,S_a)U...U G(x_b,S_b)U...UG(x_p,S_p).....U]
\label{abc}
\eea
where $a,b,...,p$ is some permutation of $12...,m$.  We now make the
second assumption of our model:

\begin{description}
\item
\ni $\underline{\mbox{\bf Assumption 2}}$

   The probabilities $f$ that plaquettes in the minimal area are pierced
by vortices are uncorrelated.  The random group orientations associated
with $S_i$  are also uncorrelated, and should be averaged.
\end{description}

   These two assumptions define our model.  They are, no doubt, an 
oversimplification of the effects of vortex thickness, but we believe
they at least provide a plausible picture of those effects.

  According to the second assumption, we are justified
in averaging independently every $G(x_a,S_a)$ over orientations in the
group manifold specified by $S_a$.  This is easily seen to give
\bea
       \overline{G}(\a) &=& \int dS \; S \exp[i\a L_3] S^\dg
                        \equiv \W_j[\a] I_{2j+1}
\non \\
  \W_j[\a] &=& {1\over 2j+1}\mbox{Tr}\exp[i\a L_3]
            =  {1\over 2j+1}\sum_{m=-j}^j \cos(\a m)
\non \\ &=& { \sin[(2j+1){\a \over 2}] \over (2j+1) \sin[{\a\over 2}] }
\label{Wj}
\eea
where $I_k$ is the $k\times k$ unit matrix.  Then
$W[C,\{x_i,S_i\}]$, averaged over all $S_i$, becomes
\beq
      W[C;\{x_i\}] = \{\prod_i \W_j[\a_C(x_i)]\} \mbox{Tr}[UU.....U]
\label{12}
\eeq
Next, take the expectation value of $\mbox{Tr}[UU.....U]$ for configurations 
$U$ with the constraint that no vortices pierce the loop, denoted $<W_0(C)>$.
Eq. \rf{12} then goes to
\beq
    <W[C;\{x_i\}]> = \{\prod_i \W_j[\a_C(x_i)]\} <W_0(C)>
\eeq
The last step is to sum over the number and position of 
vortices piercing the plane of the loop $C$, weighted by the appropriate
probability factors, and we find
\bea
      <W(C)> &=& \prod_{x} \{ (1-f) + f \W_j[\a_C(x)] \} <W_0(C)>
\non \\
        &=& \exp\left[ \sum_{x} \ln \{ (1-f) + f \W_j[\a_C(x)] \}
                \right] <W_0(C)>
\non \\
        &=& \exp[-\s(C) A] <W_0(C)>
\label{Ca}
\eea
where
\beq
       \s_C = -{1\over A}  \sum_{x} \ln \{ (1-f) + f \W_j[\a_C(x)] \}
\label{C}
\eeq
The product and sum over positions $x$ run over all plaquettes in the
plane of the loop.
The reason for not restricting $x$ to lie strictly within the minimal 
area of the loop is, again, because
the vortex core is finite.  Denote the radius of the vortex core by $R_c$.
If the center of the core lies outside the
loop, but at a distance less than $R_c$ from the 
perimeter, then it can still overlap the perimeter.  One {\it can} restrict
the sum to run over $x \in A'$, where $A'$ includes all plaquettes inside
the minimal area in the plane of the loop, as well as plaquettes in the
plane outside the perimeter, but within a distance $R_c$ of the loop. 

   Now $\s_C$ is not exactly a string tension, because it depends,
via $\a_C(x)$, on the shape (and the area) of the loop.  If, however, there
is some region where $\s_C$ changes only slowly with area, then the potential
will rise approximately linearly.  In particular, consider the limit
of very large loops.  In that case, almost every vortex which affects
the loop is entirely enclosed by the loop, and for these vortices
$\a_C(x) \approx 2\pi$.  Only those vortices near the perimeter
have $\a_C(x)$ different from $2\pi$, and as the loop becomes very large
this is a negligible fraction of the total; in particular, $A'/A \approx 1$.
This means that $\s_C$ is an area-independent constant for large loops, 
and it can be seen from eq. \rf{Wj} and \rf{Ca} that
\beq
      \s_C = \left\{ \begin{array}{cl}
                          - \ln(1-2f) & j = \mbox{half-integer} \\
                                 0    & j = \mbox{integer} 
                      \end{array}  \right.
\label{sc}
\eeq
which is the correct representation-dependence of the asymptotic string 
tension.
              
   Next, let us consider the case where $f<<1$, which is certainly
true in the lattice theory at weak coupling, and also small or medium 
size loops, where $\a_C(x)$ is also typically small.  For small $\a$,
we have from \rf{Wj}
\beq
      \W_j[\a] \approx 1 - {\a^2 \over 6} j(j+1)
\label{D}
\eeq   
Then, making an 
expansion of the logarithm in \rf{sc} and applying \rf{D},
\bea
    \s_C &=& f {1\over A} \sum_{x\in A'} \left( 1-\W_j[\a_C(x)] \right)      
\non \\
         &=&   {1\over A}\left\{ {f\over 6}\sum_{x\in A'} \a_C^2(x)  
                         \right\} j(j+1)
\eea
or just
\beq
     \s_C = {f\over 6} \overline{\a}_C^2 j(j+1)
\label{cas2}
\eeq
where $\a^2_C$ is an average value
\beq   
       \overline{\a}_C^2 =  {1\over A} \sum_{x\in A'} \a_C^2(x)
\eeq
We see that $\s_C$, for small $f$ and small loops, is proportional
to the eigenvalue of the quadratic Casimir operator.

   This result can be readily extended to any $SU(N)$ group.  In the
general case there are $N-1$ types of center vortices, corresponding
to the $N-1$ phase factors of eq. \rf{vortex}.  To the $n$-th type,
we associate probability $f_n$ to pierce a plaquette, and a group
factor
\beq
         G[x,S] = S \exp[i\va_C^n(x) \cdot \vH] S^\dg
\eeq
where the $\{H_i, ~i=1,..,N-1\}$ are the generators spanning the
Cartan sub-algebra.\footnote{It is possible that only vortices with
the smallest magnitude of center flux have substantial probability; 
i.e. $f_1=f_{N-1}$ is finite, all other $f_n$ are negligible.  This is
a dynamical issue which we cannot resolve here, and so we consider
the general case that includes all possible $f_n$.}
Following the same steps as before
\bea
       <W(C)> &=& \prod_x \left\{ 1 - \sum_{n=1}^{N-1} f_n
                        (1 - \W_r[\va_C^n(x)]) \right\}
\non \\
       \W_r[\va] &=& {1 \over d_r} \mbox{Tr} \exp[i \va \cdot \vH]
\eea
with $d_r$ the dimension of representation $r$.
Vortices of type $n$ and type $N-n$ have phase factors in \rf{vortex} 
which are complex conjugates of one another; they may be regarded as the
same type of vortex but with magnetic flux pointing in opposite directions, 
so that
\beq
       f_n = f_{N-n}  ~~~~ \mbox{and} ~~~~ 
       \W_r[\va_C^n(x)] = \W_r^*[\va_C^{N-n}(x)]
\eeq
and therefore
\bea
       <W(C)> &=& \prod_x \left\{ 1 - \sum_{n=1}^{N-1} f_n
                        (1 - \mbox{Re}\W_r[\va_C^n(x)]) \right\}
\non \\
        \s_C &=& - {1\over A} \sum_x \ln\left\{ 1 - \sum_{n=1}^{N-1} f_n
                        (1 - \mbox{Re}\W_r[\va_C^n(x)]) \right\}
\eea
Expanding the logarithm to leading order in $f_n$, expanding $\W_r[\va]$ 
to leading order in $\va$, and using the identity
\beq
 {1 \over d_r} \mbox{Tr}(H_i H_j) = {C^{(2)}_r \over N^2 -1} \d_{ij}
\eeq
one finds
\beq
      \s_C = {1\over A} \left\{ \sum_x \sum_{n=1}^{N-1} {f_n \over 2(N^2-1)}
                 \va_C^n(x) \cdot \va_C^n(x) \right\} C^{(2)}_r
\label{casN}
\eeq
where $C^{(2)}_r$ is the eigenvalue of the quadratic Casimir operator
of the $SU(N)$ group in representation $r$.

   The results \rf{cas2} and \rf{casN} might be termed 
``Casimir proportionality,'' since the non-perturbative part of the 
interquark potential, which is due to vortices,  is proportional to the 
quadratic Casimir of $SU(N)$ for small loops.  But 
this does not yet imply Casimir scaling of string tensions.  The 
parameters $\va_C^n(x)$ depend on loop size, and there is no particular reason 
to suppose that $\s_C$ is constant in the adjoint representation or, 
equivalently, that the adjoint potential is linear in some range.
Even if the adjoint potential {\it were} approximately linear in some 
interval, it is not obvious that the string tension for the fundamental 
representation, in the same range of distances, would have reached its 
asymptotic value.  To study this issue, we will return to the $SU(2)$
example.

\section{Linear Potentials and $\a_C(x)$}

   It may be possible to measure $\a_C(x)$ in computer simulations,
by the methods introduced in refs. \cite{PRD97,Zako}.  In the meantime,
it is worthwhile to ask whether there exists some reasonable ansatz for
$\a_C(x)$ which would lead to both Casimir proportionality \emph{and} linear
potentials in some region.

  To set things up, let us consider a long rectangular $R\times T$ loop
in the $x-t$ plane, with $T>>R$, in group representation $j$.  The 
time-extension $T$ is huge but fixed, so we can characterize loops $C$ 
just by the width $R$.  Let $x$ denote the $x$-coordinate of the center of 
a vortex, where it pierces the $x-t$ plane.  From the previous discussion, 
the inter-quark potential induced by vortices is easily seen to be
\beq
      V_j(R) = - \sum_{n=-\infty}^\infty \ln\{(1-f) + 
                      f\W_j[\a_R(x_n)] \}
\label{Vj}
\eeq
where $x_n = n + \oh$
(the choice of $x_n$ comes from the fact that the vortex centers lie in
the dual lattice, piercing the middle of plaquettes).
The problem is to find some reasonable ansatz for $\a_R(x)$.

   Suppose the time-like sides of the loop are at $x=0$ and $x=R$.
Then, to guide the search for an ansatz, there are a few simple conditions
that $\a_R(x)$ must satisfy:
\begin{description}
\item{1.} Vortices which pierce the plane far outside the loop don't
affect the loop.  This means that for fixed $R$, as $x \ra \pm \infty$, 
we must have $\a_R(x) \ra 0$.
\item{2.} If the vortex core is entirely contained within the loop,
then $\a_R(x)=2\pi$.  This translates as follows: 
Let $x$ be inside the loop, and $d$
be the distance from $x$ to the nearest of the time-like sides.
Then it must also be the case that $\a_R(x) \ra 2\pi$ as $d\ra \infty$. 
\item{3.} As $R\ra 0$, the percentage of any vortex core which is
contained inside the loop must also go to zero.  Thus $\a_R(x) \ra 0$
as $R \ra 0$.
\end{description}
There are an infinite number of functional forms which would meet these
conditions, but a simple 2-parameter ($a,b$) ansatz is the following:
First define
\beq
       y(x) = \left\{ \begin{array}{cl}
                     x-R & \mbox{for~} |R-x| \le |x| \cr
                     -x  & \mbox{for~} |R-x| > |x|
                   \end{array} \right.
\eeq
whose magnitude is the distance of the vortex center to the nearest
timelike side of the loop, taken negative if the vortex center is inside
the loop, and positive outside.  Then choose
\beq
      \a_R(x) = \pi \left[1 - \tanh\left(ay(x) + {b\over R}\right) \right]
\label{ansatz}
\eeq
which fulfills all three requirements.

\begin{figure}
\scalebox{.9}{\includegraphics{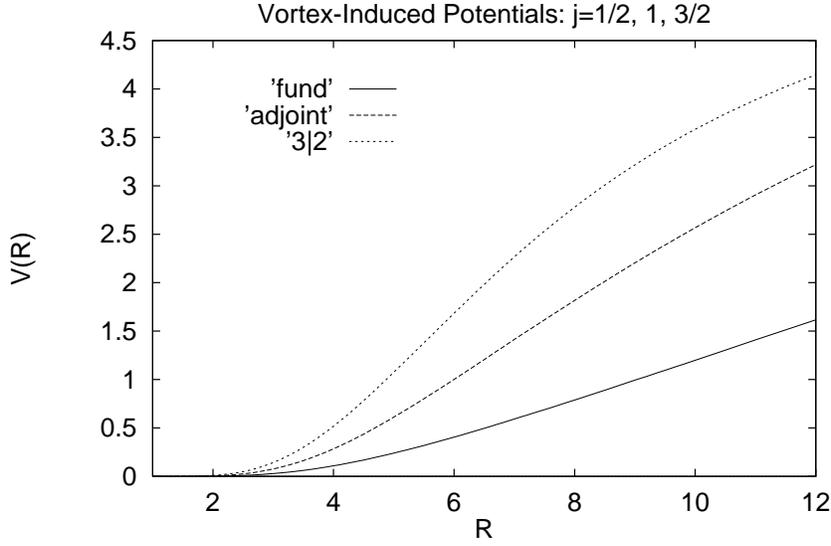}}
\caption{Interquark potential $V(R)$ induced by center vortices,
according to the model discussed in the text, for quark charges in the
in the $j=\oh,~1,~{3\over2}$ representations.}
\end{figure}

\begin{figure}
\scalebox{.9}{\includegraphics{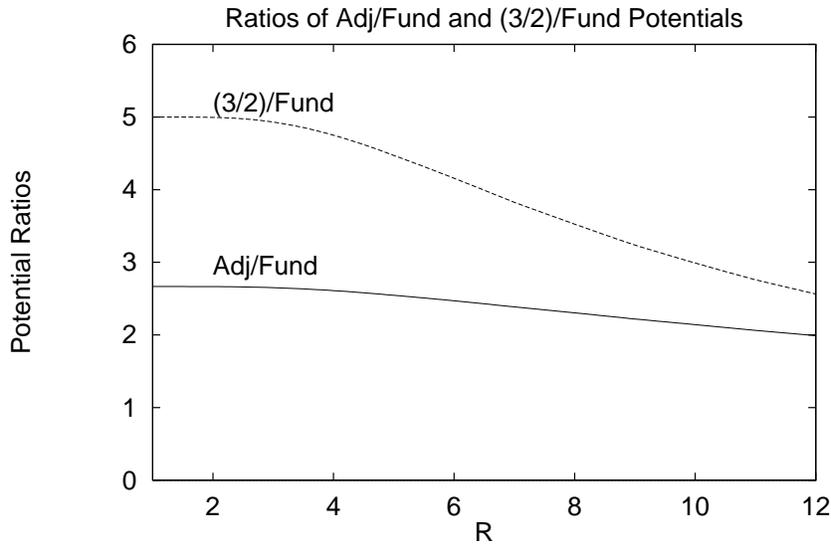}}
\caption{Ratios of $V_{3\over 2}(R)$ (upper curve) and $V_1(R)$
(lower curve) to the potential $V_{\oh}(R)$ of fundamental
representation quark charges.}
\end{figure}

\begin{figure}
\scalebox{.9}{\includegraphics{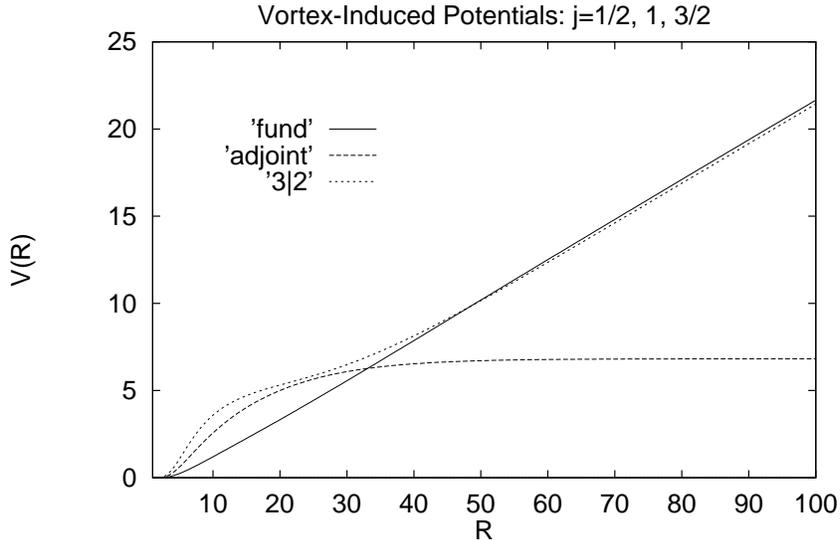}}
\caption{Same as Figure 1, 
for the range $R \in [1,100]$.}
\end{figure}

   Figure 1 shows the potentials for the $j=\oh,~1,~{3\over 2}$ 
representations, for the choice of parameters $f=0.1,~a=0.05,~b=4.$,
in the range $R \in [1,12]$.\footnote{Strictly speaking, $R$ takes on
only integer values in the lattice formulation, but we have plotted $V_j(R)$ 
over the continuous interval.}  Note that the fundamental and adjoint
potentials are roughly linear in a range from 5 or 6 to 12 lattice spacings.  
Figure 2 plots the ratios $V_1(R)/V_{1/2}(R)$, and $V_{3/2}(R)/V_{1/2}(R)$.
As expected, these ratios start out at the ratios of the corresponding 
Casimirs, $8/3$ and $5$ respectively.  It can be seen that in this interval the
adjoint/fundamental ratio drops only slowly, from $8/3$ to about $2$,
while in the same interval the ($j=3/2$)/fundamental ratio drops more
precipitously, from $5$ to about $2.5$.  Figure 3 is again a plot of all
three potentials, but this time over the range $[1,100]$.  There are
two things to notice in this last figure.  First, we see color screening 
set in as expected,
around $R=25-30$ lattice spacings.  The adjoint potential goes to a constant,
while the $j=3/2$ potential closely parallels the fundamental potential.
Secondly $-$ and this was not expected $-$ it can be seen that the
the slope of the fundamental potential, in the Casimir scaling region between
6 to 12 lattice spacings, is very close to its asymptotic value.

   We have therefore found, in this rather simple model of the vortex
core, the kind of Casimir scaling 
which is seen in Monte Carlo simulations.  A natural question is to
what extent this scaling depends on a very special choice of parameters.
The answer is that Casimir scaling, although it is not found for any
choice of parameters, is generic in a large region of the parameter
space. For example, the Casimir scaling region of Fig. 3 can be scaled
up by any factor $F$ simply by setting $a \ra a/F,~b \ra bF$. 
It is also quite easy to think up other functional forms for $\a_R(x)$
which satisy the above three conditions (e.g. 
$\a_R(x) = \b(x) - \b(x-R)$, with $\b(x) \ra \pm \pi$ in the
limits $x \ra \pm \infty$).  The possibilities are,
of course, infinite.  Some functions work better than others, but 
the existence of an approximate Casimir scaling region of some finite
extent seems to be fairly common.  The crucial ingredient is the thickness of 
the vortices, which would be on the order of $1/a$ for the choice of
$\a_R(x)$ in eq. \rf{ansatz}.  What we find
is that the thickness of the vortices must be quite large; larger, in 
fact, than the Casimir region itself, in order to see approximate 
Casimir scaling of the adjoint string tension.

\section{Conclusions}

   We have presented a scenario whereby the Casimir scaling of 
higher-representation string tensions is explained in terms of
the finite thickness of center vortices.
We do not claim to have \emph{proven} that vortex thickness is the origin 
of Casimir scaling,  but this explanation now appears to be very plausible, 
particularly if center vortices turn out to be the true QCD confinement 
mechanism.  

   Numerical tests of our scenario are in order.  If center vortices
give rise to an adjoint string tension, and if an adjoint loop is
evaluated only in those configurations where no vortex links the loop,
then the string tension should vanish.  This was found to be the case for the 
fundamental string tension, in refs. \cite{PRD97,Zako}, and should be 
testable for the adjoint string, at least in principle, by the methods 
explained in those articles.  It may also be possible to calculate 
$\a_R(x)$ from Monte Carlo simulations of fundamental loops, use that
information in eq. \rf{Vj} to derive the adjoint potential, and compare the 
derived adjoint potential with the corresponding Monte Carlo data.
  
   For every set of parameters in which we see Casimir scaling in our
simple model, the deviations from exact Casimir scaling are much
greater for the $j=3/2$ potential as compared to those of the
$j=1$ potential.  It would 
therefore be very interesting to compare Fig. 2 with the actual Monte Carlo 
data.  Calculation of the $j=3/2$ potential in the scaling region of
$SU(2)$ gauge theory, by lattice Monte Carlo techniques, is a
computationally intensive problem, but we believe it is feasible.

  Finally, there is the question raised years ago in ref. \cite{JGH}:
how can center vortices explain confinement at large $N$, where the
Casimir scaling region extends to infinity?  We have seen, in
the scenario outlined here, that the size of the Casimir scaling region
depends on the thickness of center vortices.  Already in $SU(2)$ gauge
theory, the existence of a Casimir regime implies that the 
diameter of the vortices substantially exceeds the separation length at 
which the heavy quark potential begins to grow linearly (i.e. the 
``confinement scale'').
We therefore expect that as $N$ increases, the diameter of vortices 
relative to the confinement scale will also slowly increase, probably as 
$\log(N)$.  At $N=\infty$ the vortex core would be infinite in extent, and 
the discontinuous gauge transformation exterior to the core would be pushed 
off to infinity.  It is not clear that the term ``center vortex'' remains 
a useful description of the relevant configurations in this limit.

\bigskip
\bigskip
\bigskip

\ni {\bf Note Added ~-~} We have recently
learned that a related proposal, namely, that center vortices might 
lead to a (breakable) string potential between massive gluons, was advanced 
by Cornwall \cite{Corn2} in a conference proceedings from 1983.  We thank 
Mike Cornwall for bringing this reference to our attention.

\vspace{33pt}

\newpage

\ni {\Large \bf Acknowledgements}

\bigskip

   We would like to thank J. Ambj{\o}rn and L. Del Debbio for discussions.

   This work was supported in part by Fonds zur F\"orderung der
Wissenschaftlichen Forschung P11387-PHY (M.F.), 
the U.S. Department of Energy under Grant No. DE-FG03-92ER40711
and Carlsbergfondet (J.G.), the ``Action Austria-Slovak Republic: Cooperation
in Science and Education'' (Project No. 18s41) and the Slovak Grant Agency
for Science, Grant No. 2/4111/97 (\v{S}. O.).

\end{document}